\documentclass[oneocolumn]{elsart}
\usepackage{graphicx}
\journal{}
\def\bibcode#1{(\texttt{#1})}
\begin{document}

\begin{frontmatter}
\title{On the reliability of merger-trees and the mass  growth histories of dark matter haloes}
   \author{N. Hiotelis $^{}$}
\address{1st Experimental Lyceum of Athens, Ipitou 15, Plaka, 10557, Athens,
   Greece, e-mail: hiotelis@ipta.demokritos.gr}
\thanks[now]{Present address: Roikou 17-19, Neos Kosmos, Athens, 11743
Greece}
 and
\author{A. Del Popolo $^{1,2}$}
\address{$^1$  Bo$\breve{g}azi$\c{c}i University, Physics Department,
     80815 Bebek, Istanbul, Turkey\\
$^2$ Dipartimento di Matematica, Universit\`{a} Statale di
Bergamo, via dei Caniana, 2,  24127, Bergamo, Italy\\
e-mail:antonino.delpopolo@boun.edu.tr}
\begin{small}
\begin{abstract}
      We have used merger trees realizations to study the formation of dark matter haloes. The
      construction of merger-trees is based on three different pictures about the formation of
      structures in the Universe. These pictures include: the spherical collapse (SC), the
      ellipsoidal collapse (EC) and the non-radial collapse (NR). The reliability of
      merger-trees has been examined comparing
      their predictions related to  the distribution of the number of progenitors, as well as
      the distribution of formation times, with the predictions of analytical
      relations. The comparison yields a very satisfactory agreement. Subsequently, the
      mass growth histories (MGH)  of haloes have been studied and  their formation scale
      factors have been
      derived. This derivation has been  based on two different definitions that are: (a) the scale
      factor  when the halo reaches half its present day mass and
(b) the scale factor when the mass
      growth rate falls below some specific value.  Formation scale factors follow approximately
      power laws of mass. It has also been shown that MGHs are in good agreement
with models proposed in the literature that are based on the
results of N-body simulations.  The agreement is found to be
excellent
      for small haloes but, at the early epochs of the formation of large haloes, MGHs
      seem to be
      steeper  than those predicted by the models based on N-body simulations. This
      rapid growth of mass of heavy haloes is likely to be related to a steeper central
      density profile indicated by the results of some N-body simulations.

\end{abstract}
\end{small}

\end{frontmatter}
%

\section{Introduction}

It is likely that structures in the Universe grow from small
initially Gaussian density perturbations that progressively detach
from the general expansion, reach a maximum radius and then
collapse to form bound objects. Larger haloes are formed
hierarchically by mergers between smaller ones.\\
Two different kinds of methods are widely used for the study of
the structure formation. The first one  is N-body simulations that
are able to follow the evolution of a large number of particles
under the influence of the mutual gravity from initial conditions
to the present epoch. The second one is semi-analytical methods.
Among these, Press-Schechter (PS) approach and its extensions
(EPS) are of great interest, since they allow us to compute mass
functions (Press \& Schechter  1974; Bond et al 1991), to
approximate merging histories (Lacey \& Cole 1993, LC93 hereafter,
Bower 1991, Sheth \& Lemson 1999b) and to estimate the spatial
clustering of dark matter haloes (Mo \& White 1996;
Catelan et al 1998, Sheth \& Lemson 1999a).\\
In this paper, we present merger-trees based on Monte Carlo
realizations. This approach can give significant information
regarding the process of the formation of haloes. We focus on the
mass growth histories of haloes and we compare
these with the predictions of N-body simulations.\\
This paper is organized as follows: In Sect.2 basic equations are
summarized. In Sect.3 the algorithm for the construction of
merger-trees as well as tests regarding the reliability of this
algorithm are presented. Mass-growth histories are presented in
Sect.4, while the results are summarized and discussed in Sect.5.

\section{Basic equations and merger-trees realizations}
  In an expanding universe, a
  region collapses at time $t$, if its overdensity at that time
  exceeds  some threshold. The linear extrapolation of this
  threshold up to the present time is called a barrier, B. A
  likely form of this barrier is as follows:
\setcounter{equation}{0}
\begin{equation}
  B(S,t)=\sqrt{\alpha S_*}[1+\beta(S/\alpha S_*)^{\gamma}]
  \end{equation}
  In Eq.(1)  $\alpha$, $\beta$ and $\gamma$ are constants,
   $S_*\equiv S_*(t)\equiv \delta^2_c(t)$, where $\delta_c(t)$ is the linear extrapolation
  up to the present day of the initial overdensity of  a spherically symmetric
region, that collapsed at
  time $t$. Additionally, $S\equiv \sigma^2(M)$, where $\sigma^2(M)$ is the present day  mass
  dispersion on  comoving  scale containing mass $M$. $S$ depends on the assumed
  power spectrum.
  The spherical collapse model (SC) has a barrier that does not
  depend on the mass (eg. LC93). For this model the  values of the parameters are
  $\alpha =1$ and $\beta=0$. The ellipsoidal  collapse model (EC)
  (Sheth \& Tormen 1999, ST99 hereafter) has
   a barrier that depends on the mass (moving barrier). The values of the
   parameters are $\alpha =0.707$,~$\beta=0.485$,~ $\gamma=0.615$ and are adopted
    either from the dynamics of ellipsoidal collapse or from
  fits to the results of N-body simulations. Additionally, the non-radial (NR) model
  of Del Popolo \& Gambera (1998) -that takes into account the tidal interaction
  with neighbors- corresponds to $\alpha =0.707$,~$\beta=0.375$ and $\gamma=0.585$. \\
   Sheth \& Tormen (2002) connected the
   form of the barrier with the form of the multiplicity function. They show
   that given  a mass element -that is a
  part of a halo of mass $M_0$ at time $t_0$-  the probability that at earlier
  time $t$ this mass element was a part of a smaller halo with mass $M$ is
  given by the equation:
  \begin{equation}
  f(S,t/S_0,t_0)\mathrm{d}S=\frac{1}{\sqrt{2\pi}}\frac{|T(S,t/S_0,t_0)|}{(\Delta
  S)^{3/2}}
  \exp\left[-\frac{(\Delta B)^2}{2\Delta S}\right]\mathrm{d}S
  \end{equation}
  where $\Delta S=S-S_0 $ and $\Delta B=B(S,t)-B(S_0,t_0)$ with
  $S=S(M), S_0=S(M_0)$.\\
  The function $T$ is given by:
  \begin{equation}
  T(S,t/S_0,t_0)=B(S,t)-B(S_0,t_0)+\sum_{n=1}^{5}\frac{[S_0-S]^n}{n!}
  \frac{\mathrm{\partial ^n}}{\partial S^n}B(S,t).
  \end{equation}
Setting $S_0=0$, and $B(S_0,t_0)=0$ in Eq. 3, we can predict
  the unconditional mass probability  $f(S,t)$,
  that is the probability that a mass element is  a part of a halo of mass $M$, at time
  $t$. The quantity $Sf(S,t)$
  is a function of the variable $\nu$ alone, where $\nu\equiv
  \delta_c(t)/\sigma(M)$. Since $\delta_c$ and $\sigma$ evolve
  with time in the same way, the quantity $Sf(S,t)$ is independent
  of
  time. Setting $2Sf(S,t)=\nu f(\nu)$, one obtains the so-called
  multiplicity function $f(\nu)$. The multiplicity function is the distribution
   of first crossings of  a barrier $B(\nu)$ by independent uncorrelated Brownian
    random walks (Bond et al. 1991). That is why the shape of the barrier
   influences the form of the multiplicity function. The multiplicity function
   is related to the comoving number density of haloes of mass $M$ at time $t$ -$N(M,t)$-
   by the relation,
  \begin{equation}
  \nu f(\nu)=\frac{M^2}{\rho_b(t)}N(M,t)\frac{\mathrm{d}\ln M}{\mathrm{d}\ln \nu}
  \end{equation}
  that results from the excursion set approach (Bond et al.
  1991). In Eq.4, $\rho_b(t)$ is the density of the model of the Universe at time
  $t$.\\
  Using a barrier of the form of Eq.1 in the unconditional mass probability,
   the following expression for $f(\nu)$ is found:
  \begin{equation}
  f(\nu)=\sqrt{2\alpha / \pi}[1+\beta(\alpha
  {\nu}^2)^{-\gamma}g(\gamma)]\exp\left(-0.5a\nu^2[1+\beta(\alpha \nu^2)^{-\gamma}]^2\right)
  \end{equation}
  where
  \begin{equation}
  g(\gamma)=
  \mid 1-\gamma +\frac{\gamma (\gamma
  -1)}{2!}-...-\frac{\gamma(\gamma-1)\cdot \cdot \cdot
  (\gamma-4)}{5!} \mid
\end{equation}

\begin{figure}[b]
\includegraphics[width=14cm]{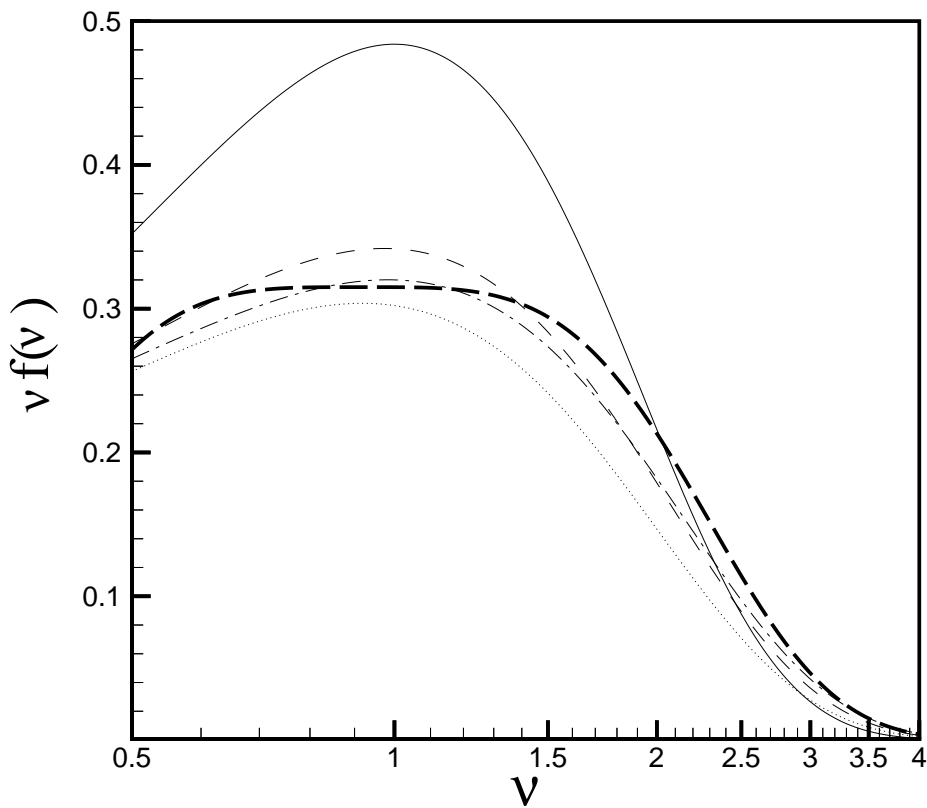}
\caption{ Various multiplicity functions. The solid line
corresponds
 to the spherical collapse model (SC) of the PS approach. The dotted
 and the dashed lines are the predictions of the Ellipsoidal collapse
 (EC) and the Non-radial collapse (NR) models, respectively. SC,
 EC and NR models are described by Eq.5 for different values of the
 parameters.  Thick dashed and the dot-dashed lines
 correspond to the predictions of J01 and ST99 models, respectively.
 These  models are described by equations 7 and 8, respectively.}\label{fig1}
 \end{figure}

  In Fig.1 we plot $\nu f(\nu)$ versus $\nu$. The solid line results from Eq. 5 for the
  values of the parameters that correspond to the SC model. The dotted line is
  derived from the values of EC model while the dashed  line corresponds to the NR model.
  For matter of completeness, we draw two more lines. The first line, the one with
  thick dashes- corresponds
  to the function of Jenkins et al. (2001, J01 hereafter). This satisfies the following equation:
  \begin{equation}
  \nu f(\nu)=0.315 exp(-\mid 0.61+ln[\sigma^{-1}(M)]\mid^{3.8})
  \end{equation}
  In order to express the above relation as a function of $\nu$, we
  substitute $\sigma^{-1}(M)=\nu/\delta_c$ and we assume a
  constant value of $\delta_c$, that of the Einstein-de Sitter Universe, namely
  $\delta_c=1.686$. The above formula is valid for $0.5 \leq \nu \leq  4.8$.\
  The second additional line, dot-dashed one, is the function proposed by ST99 and is:
  \begin{equation}
  \nu f(\nu)=A(1+\nu'^{-2p})\sqrt{2/\pi}\nu'exp(-\nu'^2/2)
  \end{equation}
  where $\nu' =\nu\sqrt{\alpha}$ and the values of the constants
   are: $A=0.322$, $p=0.3$ and $\alpha=0.707$.\\
   We have noted that the comparison of the above curves with
   the results of N-body simulations is in a very good agreement,
   except  for the SC model. Such a comparison is
   presented by  Yahagi et al. (2004). According to their results,
    the numerical multiplicity functions reside
   between the ST99 and J01 multiplicity functions at $\nu \geq 3$ and are below the ST99
   function at $\nu \leq 1$. Additionally, the numerical multiplicity functions have
   an apparent peak at $\nu \sim 1$ -as those given by Eq. 5- instead of the plateau
   that is seen in the J01 function.\\

\section{The construction of merger-trees}
  Let there be a number of haloes with the same present day mass $M_0$.
  The purpose of merger-trees realizations is to study the past of these haloes.
  This is done by finding the distribution of their progenitors (haloes that merged
  and formed the present day haloes) at previous times. A way to do this is by
  using  Eq.2.  First of all, one has to use a model for the Universe and
  a
  power spectrum. In what follows we have assumed a flat model for the Universe with
  present day density parameters $\Omega_{m,0}=0.3$ and
   $ \Omega_{\Lambda,0}\equiv \Lambda/3H_0^2=0.7$ where
  $\Lambda$ is the cosmological constant and $H_0$ is the present day value of Hubble's
  constant. We used the value $H_0=100~\mathrm{hKMs^{-1}Mpc^{-1}}$
  with $h=0.7$ and a system of units with $m_{unit}=10^{12}M_{\odot}h^{-1}$,
  $r_{unit}=1h^{-1}\mathrm{Mpc}$ and a gravitational constant $ G=1$. At this system of units
  $H_0/H_{unit}=1.5276.$\\
 Regarding  the power spectrum- that defines the relation between
  $S$ and $M$ in Eq.2- we  employed the $\Lambda CDM$ formula proposed by
  Smith et al. (1998). The power spectrum is smoothed using the top-hat window function and
  is normalized for $\sigma_8\equiv\sigma(R=8h^{-1}\mathrm{Mpc})=1$.\\
For each one of the SC, EC and NR models, we studied four
different cases.  These four cases differ to the present day mass
of the haloes under study and are denoted as SC1, SC2,
   SC3,  SC4 for the SC model, EC1, EC2, EC3 and EC4 for the EC
   model and NR1, NR2, NR3 and NR4 for the NR model.
 The first case of each model (SC1, EC1 and NR1) corresponds to haloes with
 mass 0.1 -measured in our system of units- while the second case (SC2, EC2 and NR2)
    corresponds to mass 1. SC3, EC3, NR3 correspond to haloes with mass 10
  and SC4, EC4, NR4 correspond to masses 100.\\
  Pioneered works for the construction of merger-trees are those of
  LC93,  Somerville \& Kollat (1999), ST99 and van de   Bosch (2002, vdB02).
   Many of their ideas
  are used for the construction of our algorithm, that is described below.\\
   First, let us define as $N_{res}$, the number of realizations used. This
   number is  the total number of present day haloes of given mass $M_0$.
  A mass cutoff $M_{min}$ is used, that is a lower limit for the mass of
   progenitors. No progenitors with mass less than $M_{min}$ exist in the merger
   history of a halo.  $M_{min}$ is set to a fixed fraction of $M_0$. In what follows,
    $M_{min}=
   0.05M_{0}$ (see also vdB02). Also, $a_{min}$ is the minimum value of the
   scale factor. We set $a_{min}=0.1$.  Merging histories do not extend to values of $a <
   a_{min}$. A  discussion about the choice of $a_{min}$ is
   given in Section 4.\\
     Useful matrices which have been used are the following: $M_{par}$ (the masses of the
   parent haloes),~$M_l$ (the masses that are left to be resolved) and $M_{pro}$ (the masses of the
   progenitors). The  argument $i$
   takes the values from $1$ to $N_{max}$, where $N_{max}$ is the number
   of haloes that are going to be resolved. Initially,
   $N_{max}=N_{res}$. The values of $\Delta\omega,~N_{res},~M_{min},~a_{min}$
    are used as input. We discuss the choice of the value of  $\Delta\omega$ in Section 4.
    The scale factor $a$   is set to its initial value, (that is the present day
   one),   $a=a_0=1$. Then, the  matrices $M_{par}$ and $M_l$ take their values:
   $M_{par}(i)=M_0$,~ $M_{l}(i)=M_0$ for $i=1,N_{res}$.  It is convenient to set
   a counter for the
   number of time steps, so we set $I_{step}=0$. The rest of
   the procedure consists of the following steps:\\
   \textbf{1st step}: $I_{step}=I_{step}+1$.\\
   The equation $\delta_c(a_p)=\Delta \omega +\delta_c(a)$
   is solved for $a_p$, that is the new, (current) value of the scale factor. \\
   \textbf{2nd step}: for $i=1, N_{max}$\\
   \textbf{3rd step} A value of $\Delta S$ is chosen from the desired distribution.\\
   \textbf{4th step}:
    The mass $M_p$ of the progenitor is found, solving for $M_p$ the equation:
   $\Delta S=S(M_{p})-S(M_{par}(i))$. If $M_{p}\geq M_l(i)$  then, we
   return
   to the 3rd step. Else, the halo with mass $M_p$ is a progenitor of the parent
   halo $i$ . The mass to be resolved is now given by:  $M_l(i)=M_l(i)-M_p$.
    If $M_p< M_{min}$ then $M_p$ is too small and
   is not considered as a real progenitor, therefore we return to the 3rd  step.
   Else,\\
   \textbf{5th step}:
   The number of progenitors is increased by one, ($N_{pro}=N_{pro}+1$), and the new
   progenitor is stored to the list $M_{pro}(N_{pro})=M_p$.\\
   \textbf{6th step}: If the mass left to be resolved exceeds the minimum mass,
   ($M_l(i)>M_{min}$), we go back to the 3rd step. Else, we return to the
   2nd step, where the next value of $i$ is treated.\\
   \textbf{7th step}:  We set the new number of haloes $N_{max}$ to be the
   number of progenitors, $N_{max}=N_{pro}$. We store the progenitors,
   $M_{pro}(j),~ j=1,N_{max}$.  The masses of new parents are
   defined as the masses of progenitors by $M_{par}(j)=M_{pro}(j),j=1,N_{max}$.
   The value of the scale factor is updated $a=a_p$, as well as
   the mass left to be resolved, $M_l(j)=M_{par}(j),i=j,N_{max}$. If $a\leq a_{min}$ we stop.
   Else, we return to the first step.\\
     Output is the list of progenitors after a desired number of time
   steps. Regarding the third step, we have the following
   remarks. For the SC case, the change of variables
   $x \equiv \frac{\delta_c(t)-\delta_c(t_0)}{\sqrt{S(M)-S(M_0)}}$ leads to a
   Gaussian distribution with zero mean value and unit variance for the new variable $x$.
   So, we pick values $x$ from the above Gaussian (that is the well-known and very fast
   procedure Press et al. 1990). The values of $\Delta S=x\Delta \omega$ are distributed
   according to the distribution described in Eq. 2. The
   distributions in the EC and NR  cases cannot be expressed using a
   Gaussian, so a more general method is used. This method,
    described in Press et al. (1990) is as follows:\\
   Let's suppose that we would like to pick numbers from a distribution function
    $G(x)$, where the variable
   $x$ takes positive values in the interval $[0,c]$. First, we calculate the integral
   $\int_{0}^c G(t)\mathrm{d}t=A$. We note that
   $A$ is not necessarily equal to unity. Then, a number, let say $y$,
   between zero and $A$ is picked from a uniform distribution. The integral
   equation:
   \begin{equation}
   \int_0^xG(t)\mathrm{d}t=y
   \end{equation}
   is solved for $x$. The resulting points $(x,G(x))$ are then
   uniformly distributed in the area between the graph of $G$,
   the x-axis and the lines $x=0$ and $x=c$. The values of $x$
   have the desired distribution. We have to note that this
   procedure is -as it is expected- more time demanding than the
   SC. This is mainly due to the numerical solution of the
   integral equation. However, it has the advantage of being
   general.\\
   In Eq.2, $f$ becomes an one variable function for given values of
   $S_0,~t_0,$ and $t$. Then, the
   above described general procedure is used in order to pick
   values of $S$. For numerical reasons, it is convenient to use a new
   variable $w=1/\sqrt{S-S_0}$.\\
   The distribution of the masses of progenitors is found using the following procedure:
   Let $N_p$ be the number of progenitors of $N_{res}$ parent haloes
   at time $t$  and let $m_{min},~ m_{max}$ be the minimum and the maximum mass
    of these progenitors,
   respectively. We divide the interval $[m_{min},
   m_{max}]$ to a set of $k_{max}$ intervals of length $\Delta m$. Then, the
   number of progenitors, $N_{pro,k}$, in the interval $ [m_{min}+k\Delta m,~m_{min}+(k+1)\Delta
   m]$ for $k=0,k_{max}-1$ is found. The distribution of progenitors masses is then given
   by $ N_{m-tree}\equiv \frac{N_{pro,k}}{\Delta m N_{res}}$.
   This distribution has to be compared with the
   distribution that results from the analytical relations. Thus, multiplying
Eq.2  by $M_0/M $, one finds the expected number of progenitors at
$t$ that lie
   in the range of masses $M,~M+\mathrm{d}M$. This number obeys the equation:
  \begin{equation}
  N(M,t/M_0,t_0)\mathrm{d}M=\frac{M_0}{M}f(S,t/S_0, t_0)\mid\frac{\mathrm{d}S}{\mathrm{d}M}\mid\mathrm{d}M
  \end{equation}
  which is known as the number-weighted probability for $M$.
\begin{figure}[b]
 \includegraphics[width=14cm]{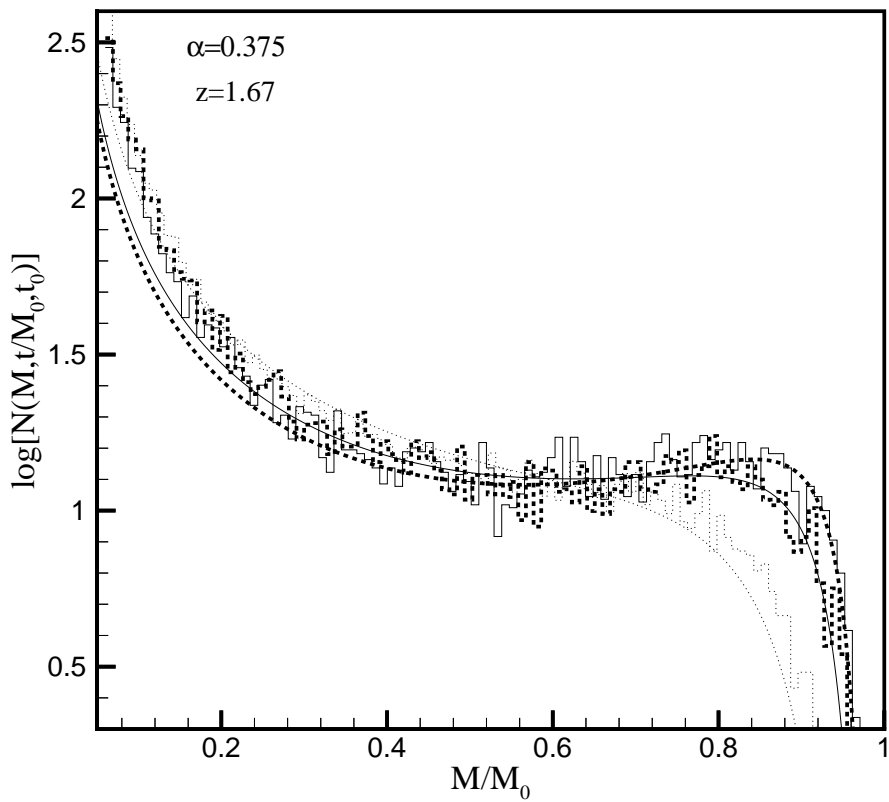}
\caption{ The distribution of progenitors for various
 models used at scale factor $a=0.375$ (redshift
$z=1.67$). All curves correspond to a halo with present day mass
 $M_0=0.1$. The smooth  dotted and the crooked dotted lines give $log(N)$ and $log(N_{m-tree})$
 versus $M/M_0$ for the SC1 model, respectively. Smooth solid and crooked solid lines
 give $log(N)$ and $log(N_{m-tree})$ for the NR1 model, respectively.
 In a similar way, the line with short thick dashes correspond to EC1
 model.}\label{fig2}
 \end{figure}
  In Figs 2 and 3, the distributions of progenitors are plotted for all models studied, for
  two different values of the scale factor.
  The results of Fig.2 correspond to scale factor
  $a=0.375$ and those of Fig.3 to $a=0.2$. Details are given in
  the captions of the Figs. Large differences
  between the predictions of spherical and non-spherical models
  are shown. These differences are very clear and they can be seen even if
  a relatively  small sample of haloes is used (we used $N_{res}=5000$). On the
  other hand, the predictions for the two non-spherical models NR and EC are very close.\\
 \begin{figure}[b]
 \includegraphics[width=14cm]{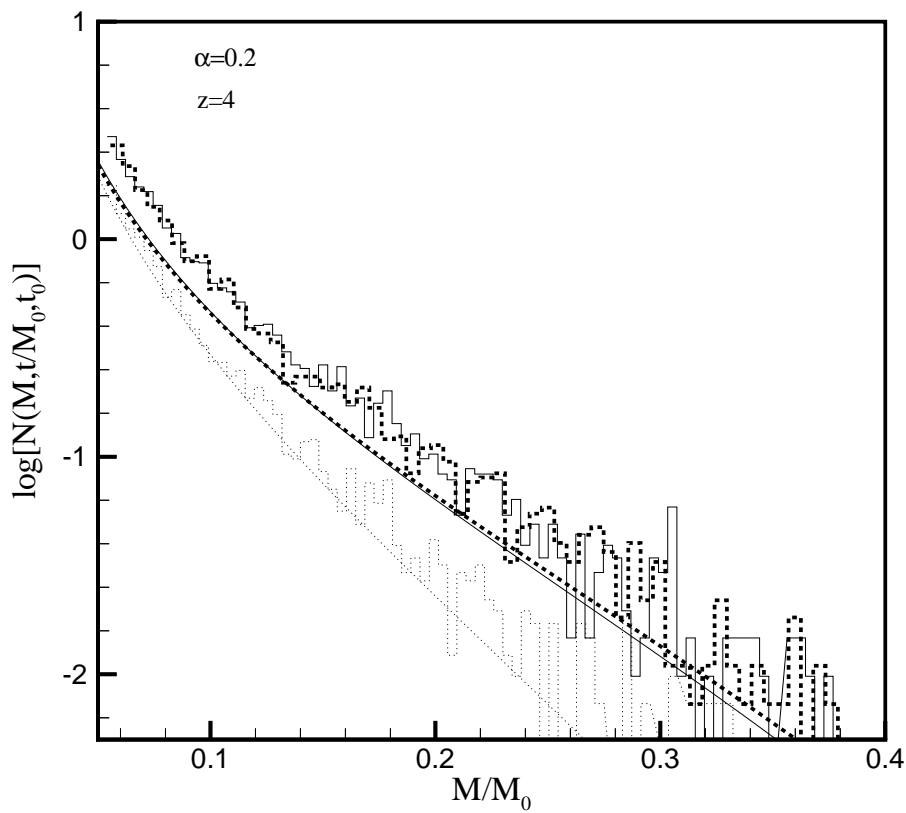}
 \caption{The distribution of progenitors for various
 models studied at scale factor $a=0.2$ (redshift $z=4$). All curves correspond to a halo with present day mass
 $M_0=10$. The smooth  dotted and the crooked dotted lines give $log(N)$ and $log(N_{m-tree})$
 versus $M/M_0$ for the SC3 model, respectively. Smooth solid and crooked solid lines
 give $log(N)$ and $log(N_{m-tree})$ for the NR3 model, respectively.
 Lines with short thick dashes correspond to the results of EC3 model. The smooth line
 gives $log(N)$ while the crooked one gives $log(N_{m-tree})$.}\label{fig3}
 \end{figure}
The time evolution of halo's mass is known as its mass growth
history. However, a clear
  definition of the halo's mass at time $t$ is needed. One can define as halo's mass at
a step, the mass of its most massive progenitor at that step.
   We call this procedure the
  \textit{most massive} progenitor approximation.
  A different approximation is the one followed by vdB02. We call
  this approximation
  the  \textit{main} progenitor approximation and it is as follows:
  At the first step, the most massive progenitor $M_{mmp}$
  of a halo is found. This is the halo's mass at that time.
  At the next step, the most massive progenitor of the halo with mass $M_{mmp}$ is
  found and is considered as the new halo's mass. The procedure is  repeated for
  the next steps. According to the definition of LC93, the scale factor at which
  the mass of the \textit{main} progenitor equals half  the present day mass
  of the halo is called the \textit{formation} scale factor. We denote it
  by $a_f$. A detailed description of the procedure follows:\\
     \textbf{1st step}: $I_{step}=I_{step}+1$.\\
   The equation $\delta_c(a_p)=\Delta \omega +\delta_c(a)$
   is solved for $a_p$ that is the new (current) value of the scale factor. \\
   \textbf{2nd step}:for $i=1,N_{max}$\\
   The mass of the most massive progenitor of parent $i$ is set equal to
   zero ($M_{mmp}(i)=0$).\\
   \textbf{3rd}. A value of $\Delta S$ from the desired distribution is chosen.\\
    \textbf{4th step}:
    A mass $M_p$ is found, solving for $M_p$ the equation:
   $\Delta S=S(M_{p})-S(M_{par}(i))$. If $M_{p}\geq M_l(i)$  then, we go
   back to the 3rd step. Else, the halo with mass $M_p$ is a progenitor. We set
   $M_l(i)=M_l(i)-M_p$. If $M_p< M_{min}$ then $M_p$ is too small and
   is not considered as a real progenitor and so we return to the 3rd step.
    Else,\\
    \textbf{5th step}:
    The most massive progenitor at this time step is found.\\
    $M_{mmp}(i)=max(M_p,M_{mmp}(i))$.\\
    \textbf{6th step}: If  the mass left to be resolved exceeds the minimum mass,
   ($M_l(i)>M_{min}$), we go back to the third step. Else, we check if the mass of
   the most massive progenitor is for the first time smaller than half the initial mass
   of the halo. If this is correct, then the formation scale factor of this halo is defined
   by linear interpolation:
   \begin{equation}
   a_{form}(i)=a_p+\frac{0.5M_0-M_{mmp}(i)}{M_{par}(i)-M_{mmp}(i)}(a-a_p)
   \end{equation}
   Otherwise, we proceed with the next $i$.\\
   \textbf{7th step}:
   The list of the most massive progenitors is used, in order to find the
   mean mass at this time step.
   The time of the scale factor is updated $a=a_p$, as well as
   the mass left to be resolved, $M_l(j)=M_{par}(j),~j=1,N_{res}$. If $a\leq
   a_{min}$ we  stop, else we go to the first step.\\
   We have to note that the halo's mass at a step -derived by the
   above procedure- does not give the mass of its most massive progenitor at
   that step. For example, let a halo of mass $M_0$ that has at the first step two
   progenitors with masses $M_1$ and $M_2$ with $M_1 > M_2$. Then, the mass of the halo
   at the second step is defined as the mass of the most massive progenitor ($M_1$).
   This progenitor is not necessarily the most massive of all progenitors at that step, since
   it could be less massive than one of the progenitors of $M_2$.
   The procedure of \textit{main} progenitor described above has
   the advantage of being economic. It does not  require the construction of a complete set
   of progenitors in order to find  the most massive one at a
   step.\\
   We tested both the \textit{most massive}  and the \textit{main}
   progenitor approximations and we found small differences only
   for massive haloes. In order to be able to compare our results with those
   of other authors, we used the \textit{main} progenitor
   approximation.  \\
   A test for the reliability of merger-trees is the comparison
   of the distribution of formation scale factors with the one given
 by the analytical relations
   that follow.
   The  probability that the mass of the progenitor, at time $t$, is larger than
  $M_0/2$ is given by:
  \begin{equation}
  P(t,M_p>M_0/2)=\int_{M_0/2}^{M_0}N(M,t/M_0,t_0)\mathrm{d}M
  \end{equation}
\begin{figure}[b]
 \includegraphics[width=14cm]{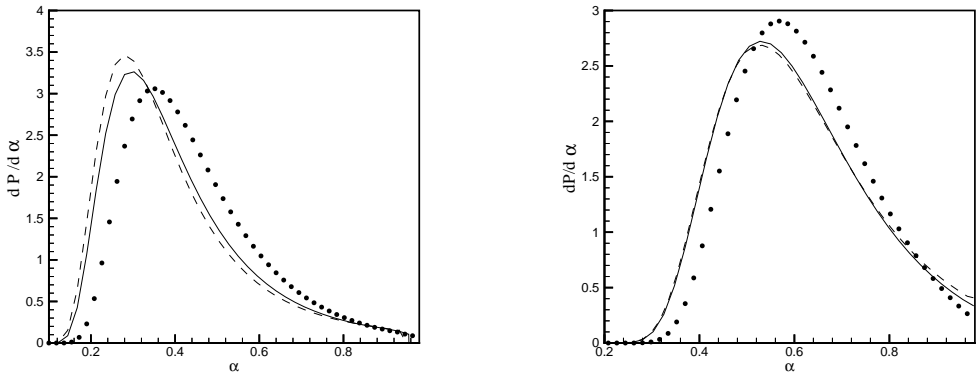}
 \caption{ The distribution of formation scale factors  as it results from Eq.13 for the
 three models used in this paper. The dotted lines
 correspond to the SC model, the solid ones to the NR model and the dashed to the EC model.
 The left snapshot corresponds to haloes with present day mass 0.1 -in our system of units-
 while the right snapshot to haloes with present day mass 100.}\label{fig4}
 \end{figure}
  The formation time is defined (LC93) as the epoch when the mass
  of the halo, as it grows by mergers with other haloes, crosses
  the half of its present day value.
   Then, $P(t,M_p>M_0/2)$ is the probability the halo with present day mass
  $M_0$ had a  progenitor heavier than $M_0/2$ at $t$, which is equivalent
  to the probability
   that the  halo is formed earlier than $t$, $P(<t,M_0)$. Thus, $P(t,M_p>M_0/2)=P(<t,M_0)$.
  Differentiating with respect to $t$, we obtain:
  \begin{equation}
  \frac{\mathrm{d}P(<t,M_0)}{\mathrm{d}t}=\int_{M_0/2}^{M_0}\frac{\partial}{\partial t}
  \left[N(M,t/M_0, t_0)\right]\mathrm{d}M
  \end{equation}
  which gives the distribution of formation times.\\
  The distribution of formation scale factors as it results from Eq.13 is
  plotted in Fig. 4. The left hand side snapshot corresponds to haloes with present day
   mass 0.1, while  the   right hand side one to haloes with present day mass 100.
   In every snapshot dotted, solid and dashed
  lines correspond to  SC, NR and EC models, respectively.
  Notice that the distributions that correspond to NR and EC models are shifted to the left
  relative to the SC model.  This means that structures formed earlier in the NR and EC
  models.
As it can be seen in  Fig. 4, the differences between spherical
and non-spherical models are smaller for
  larger haloes. As it is known in the literature(e.g Yahagi et
al, 2004 and references therein), the results of non-spherical
models are in good agreement with the results of N-body
simulations. Consequently, for massive haloes, the results of
N-body simulations are close to the predictions of the spherical
models, as well.
 This disagrees with the results of vdB02 which predicts that the
  disagreement between spherical model and N-body simulations is larger for
  massive haloes.\\
\begin{figure}[t]
 \includegraphics[width=14cm]{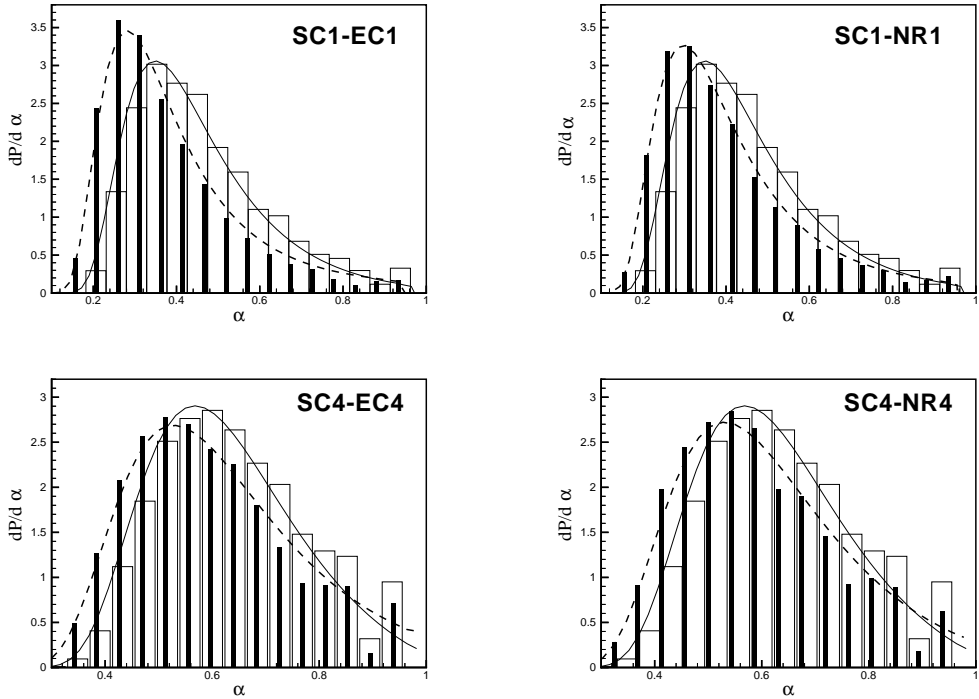}
 \caption{ Distributions of formation scale factors derived either from analytical relations
  (Eq.13) or from the analysis of merger-trees  results (see
  Eq.11).} \label{fig5}
 \end{figure}
Fig.5 consists of four snapshots. Bars show the distributions of
formation
  scale factor as they result from Eq.11. Wide bars are the
  predictions of SC model, while narrow  filled bars are the
  results of EC or NR models. Solid lines are the predictions of Eq.
  13  for the SC model, while dashed lines are the predictions of the same
  Eq. for EC or NR models. The two snapshots of the first row
  show the results of SC1 and EC1 (left snapshot) and SC1 and NR1 (right
  snapshot) while the two snapshots of the second row show  the results
  of SC4 and EC4 (left snapshots)  and SC4 and NR4 (right snapshot). It is shown that
  the predictions of Eq. 13 are in very good agreement with the numerical distributions.
  Additionally, it is clear that the differences between the distributions of various models,
   resulting from Eq. 13,  are successfully  reproduced by numerical
   distributions. It is also shown that differences between spherical and non-spherical models
   are more significant for haloes of small mass.

\section{Mass-growth histories}
We define as mass-growth history, MGH, of a halo with present-day
mass $M_0$,  the curve that shows the evolution of
$\widetilde{M}(a)\equiv <M(a)>/M_0$ where $<M(a)>$ is the mean
mass at scale factor $a$ of  haloes  with present day mass $M_0$.
Two input parameters are used in the method of construction
merger-trees in the procedure described in the previous section.
The first one is the  'time-step' $\Delta\omega$. For values of
$\Delta\omega \leq 0.3$, it is shown (vdB02) that  MGHs are not
time-step dependent. According to that, we used a value
$\Delta\omega =0.1$. The second parameter is $a_{min}$.  Our
results are derived for $a_{min}=0.1$. In order to justify this
choice, let us discuss the following points regarding the
definition of the mean mass. Let $N_h(a)$ be the number of haloes
at scale factor $a$ that have masses greater than $M_{min}$.
Obviously, $N_h(1)=N_{res}$. Let also $M_{h,i}(a)$ be the mass of
the $i^{th}$ halo at scale factor $a$.  We define by $M_k(a)$ the
sum $\sum M_{h,i}(a)$, where for $k=1$ the sum is extended to all
haloes that satisfy $ M_{h,i}(a)\geq M_{min}$. For $k=2$ the sum
is extended to all haloes (obviously in that case, the masses of a
number of haloes are equal to zero since, by the construction of
merger-trees, their mass histories are not followed beyond
$M_{min})$. Finally, for $k=3$ the sum is over those haloes  that
satisfy $M_{h,i}(a_{min})\geq M_{min}$. Then, we can define the
mean mass at $a$ in three different ways: First
$M^{mean}_1=M_1(a)/N_h(a)$, second $M^{mean}_2=M_2(a)/N_{res}$ and
third $M^{mean}_3=M_3(a)/N_h(a_{min})$.\\
As $a$ evolves from its present day value to smaller values, the
number $N_h(a)$ becomes smaller and the three mean values defined
above become different. It is clear that $M^{mean}_1$
overestimates the mean value, while $ M^{mean}_2$ underestimates
it. On the other hand,  $ M^{mean}_3$ requires a large number of
$N_{res}$ so that  the remaining sample of haloes $N_h(a_{min})$
to be large enough. The choice of the value $a_{min}=0.1$ is
justified by the fact that for $a\geq a_{min}$ the differences in
the above three
definitions are negligible.\\
The following results have been derived using a number of 10000
haloes. We found that a number of haloes greater than 1000 is
sufficient for producing smooth MGHs that are in exact agreement
 with the ones of our results.\\
Fig.6 consists of three snapshots. The first snapshot shows the
results of the SC model. The other two snapshots correspond to the
EC and to the NR model, respectively. Every snapshot contains four
lines. From top to bottom these lines correspond to the cases 1,
2, 3 and 4, respectively. It is clear that -at present time- the
slopes  that correspond to smaller masses are smaller. This
sequence of slopes is a sequence of  formation times. Large haloes
continue to increase  their masses with a significant rate, even
at the present epoch, while the mass growth of small haloes is
negligible. Thus, small haloes have been formed earlier. However,
a \textit{formation} scale factor can be defined by the condition
that the growth rate of mass falls below some value.   Bullock et
al. (2001, B01) used as a characteristic scale factor, the one
satisfying the relation
$\mathrm{d}\log\widetilde{M}(a)/\mathrm{d}\log(a)=\mu$,
 where $\mu$ is a
specific value, typically equal to 2. We used the value $\mu =2$
and  we denote the solution of the equation $a_c$. Additionally,
we denote $a_f$ the solution of the Eq. $\widetilde{M}(a)=1/2$ and
we plot the results in Fig.7. This figure consists of two
snapshots.
\begin{figure}[t]
 \includegraphics[width=14cm]{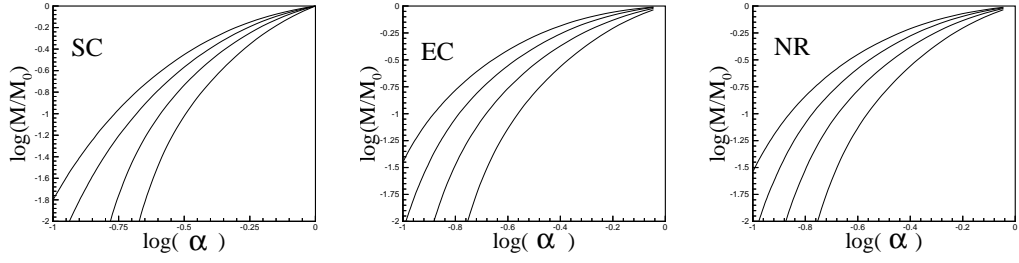}
 \caption{Mass growth histories for the three models studied. Lines - from
 top to bottom- correspond to cases 1, 2, 3 and 4 respectively.
 The growth rate of smaller haloes falls below some critical
  value earlier than that of larger ones. This means that smaller haloes are formed earlier}
  \label{fig6}
 \end{figure}
\begin{figure}[b]
 \includegraphics[width=14cm]{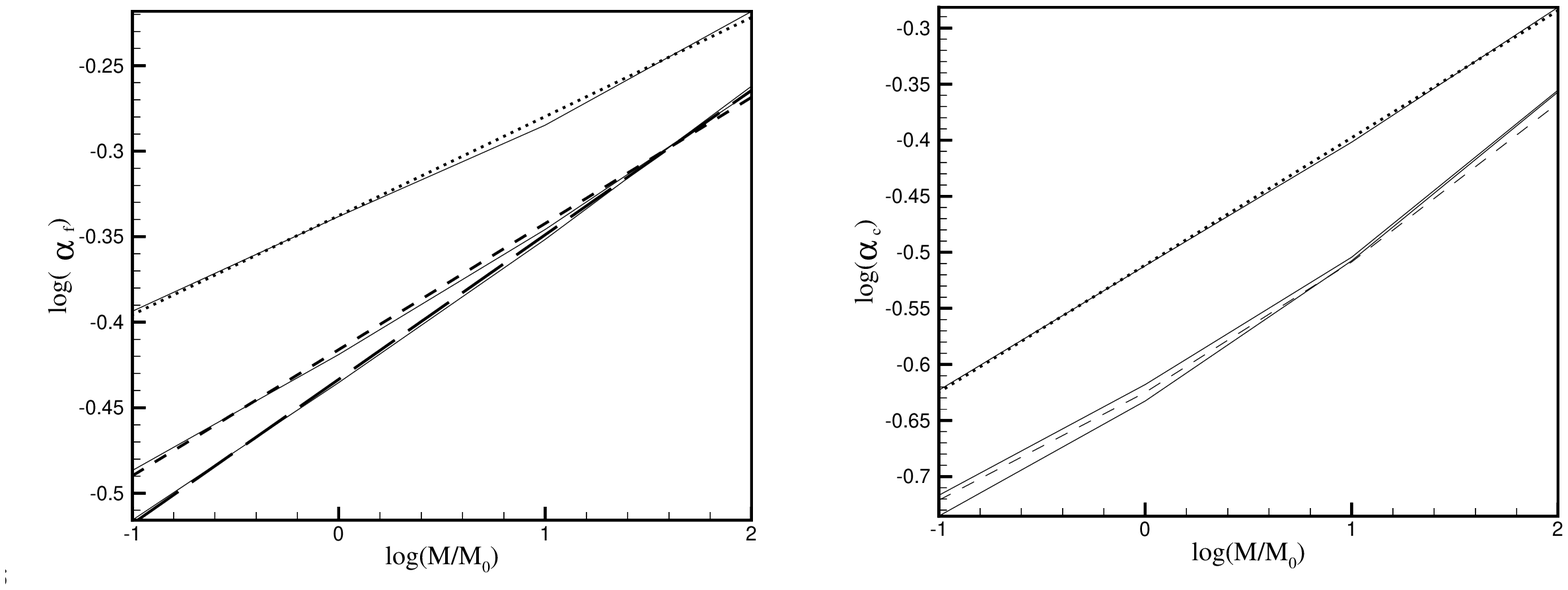}
 \caption{The relation between the 'formation' scale factor and
 mass. The left snapshot shows the relation between $\alpha_ f$ and $M$. From top to bottom,
 the solid  lines are the plots of $log(a _f)$  versus $log(M/M_0)$ for
 the SC, NR and EC models, respectively. The dotted, short-dashed and long-dashed lines are
 the linear fits of the above lines, respectively.  From top to  bottom in the right
snapshot, the solid lines  are the plots of $log(a _c)$ versus
$log(M/M_0)$ for
 the SC, NR and EC models, respectively. The dotted line is the linear fit of the SC model,
  while the dashed line is a prediction of $a _c$ given by the toy-model of W02
 (see text for more details)}\label{fig7}
 \end{figure}
 Left snapshot shows $log(a_f)$ versus $log(M/M_0)$ for the three
 models studied. From top to  bottom,  solid lines show
 the predictions of SC, NR and EC models, respectively. These lines
 are well fitted by linear approximations that are shown by
 dotted, dashed and long-dashed lines, respectively. Thus $a_f$
 and $M$ are related by a power law $a_f(M)\sim M^n$. The values
 of $n$ are $n=0.058$, $n=0.074$ and $n=0.085$ for SC, NR and EC
 models, respectively. The solid lines in the right snapshot show
 $log(a_c)$ versus $log(M/M_0)$. The dotted line is the linear
 approximation of the results of SC model. It can be seen that this
 approximation is very satisfactory. Linear approximations are not
 satisfactory for non-spherical models. The dashed line in this snapshot shows the
 predictions of the toy-model of Bullock et al. (2001, B01 hereafter) that calculates
 $a_c$ from the power spectrum using following procedure:
 Let a halo with mass $M$. The scale factor at the epoch of its collapse $a_c$, is
 found by the solution of equation.:
 \begin{equation}
 \sigma[M_*(a_c)]=\sigma(FM)
 \end{equation}
 where, $M_*$ is the typical collapsing mass satisfying:
 $\sigma[M_*(a)]=1.686D(1)/D(a)$, and $F$ is a constant. $D(a)$ is
the growth factor predicted by linear theory (Peebles, 1980).
  The dashed line presented in the right snapshot of  Fig.7 is  the prediction
  of the above procedure for $F=0.004$ and it is a good approximation of our results.\\
  A less accurate but very practical approximation of $a_c$ is given (see also Suto 2003) by
  \begin{equation}
  a_c(M)\approx0.23M^{0.13}
  \end{equation}
  where, $M$ is expressed in our system of units.  A value of about $0.13$ for the
  exponent of the above relation is confirmed by our results too, since  the best
  linear fits of the solid lines in the right
  snapshot of the Fig.7, are of the form $a_c(M)\sim M^n$ with
  $n=0.1135$, $n=0.12$ and $n=0.1256$ for the SC, NR and EC
  models,
  respectively.\\
 In a complementary paper of that of B01, Wechsler et al. (2002, W02
 hereafter), based on the results of their N-body simulations,
  proposed for the growth of mass of dark matter haloes the exponential  model:
  \begin{equation}
  \widetilde{M}(a)=\exp[-a_c \mu (1/a-1)]
  \end{equation}
  In  Fig.8,  we compare our MGHs with the predictions of the model of W02.
  This   figure   consists of twelve snapshots.
   Every row has four snapshots that correspond to the
  four cases of masses (0.1, 1, 10  and 100) studied for every model. The
  predictions of SC model are presented in the first row. The second row shows  the
  predictions of NR and the third one the predictions of EC model. The solid lines show our
  MGHs while the dashed lines are the predictions of the model of W02 given by Eq.16. The
  values of $a_c$ used are those shown by the solid lines of Fig.7 that
  are predicted by the numerical solution of the equation
  $\mathrm{d}\log(\widetilde{M})/\mathrm{d}\log(a)=2.$ Thus, no other information (such
  as the cosmology, power spectrum etc) is used but only the MGHs of
  haloes. All snapshots refer to $\widetilde{M}(a)\geq 0.01$.\\
  The conclusion from  Fig.8 is clear. The predictions of Eq.16
  are very good approximations to the MGHs predicted by  merger-trees realizations.
  \begin{figure}[b]
  \includegraphics[width=14cm]{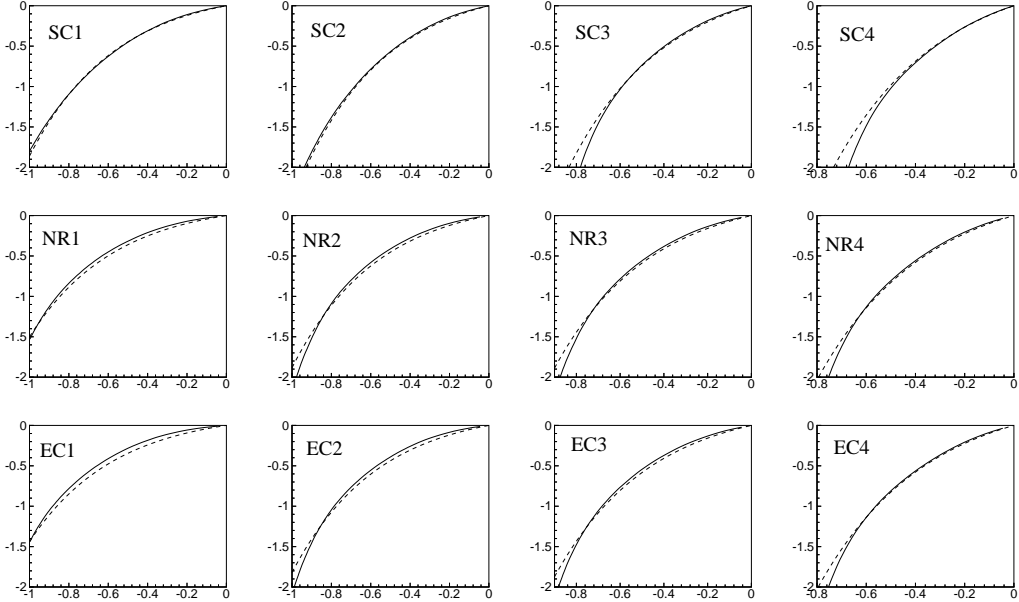}
  \caption{ $\log(M/M_0)$ versus $\log(a)$ for all the models studied. The solid lines are the
   predictions of merger-trees realizations while the dashed ones are their fits by the model
   given in Eq.16 (see text for more details)}
  \label{fig8}
  \end{figure}
 The quality of fit depends on the mass of the halo and not on the model used. For
 example,
   the left column of Fig. 8 shows a very good fit for all haloes with mass $0.1$. A more careful
   look at this figure shows that the fit becomes less satisfactory for the earlier stages of
   the evolution of large haloes. In fact, the predictions of
   merger-trees are steeper curves than those of the model of W02.
   This can be seen clearly at the 3thd and 4th column that show
   the predictions for haloes with masses $10$ and $100$.
   We note that the mass growth rate can be connected - under some
   assumptions- with the density profile of the halo (e.g.  Nusser \& Sheth 1999,
   Manrique et al. 2002, Hiotelis 2003).  According to the 'stable-clustering' model used by
   the above authors, haloes grow inside-out. This means that the accreted mass is deposited
   at an outer spherical shell without changing the inner density
   profile. We recall that the radial extent of a halo is defined by its virial
    radius $R_{vir}$, that is the radius that contains a mass
    with mean density $\Delta_{vir}$ times the mean density of the Universe
    $\rho_b$.  $\Delta_{vir}$ is, in general, a function of the scale factor $a$ but in
    many applications it is set equal to a constant value of $200$ (that is a value
    for $\Delta_{vir}$ for an Einstein de Sitter Universe).
    The mass $M_{vir}$ contained inside $R_{vir}$ at scale factor $a$ satisfies
    the two following equations:

    \begin{equation}
    M_{vir}(a)=4\pi\int_0^{R_{vir}(a)}r^2\rho(r)\mathrm{d}r
    \end{equation}
    \begin{equation}
    M_{vir}(a)=\frac{4}{3}\pi \Delta_{vir}(a)\rho_b(a)R^3_{vir}(a)
    \end{equation}

    If the mass history of the halo is known, (that is the evolution of its mass as a function
    of the scale factor), then the density profile can be calculated by differentiating Eq. 17
     with respect to $a$, which gives:
     \begin{equation}
     \rho(a)=\frac{1}{4\pi
     R^2_{vir}(a)}\frac{\mathrm{d}M_{vir}(a)}{\mathrm{d}a}
     \left(\frac{\mathrm{d}R_{vir}(a)}{\mathrm{d}a}\right)^{-1}
     \end{equation}
     where $\mathrm{d}R_{vir}(a)/\mathrm{d}a$ is
     calculated by differentiating Eq.18. Finally Eqs. 19 and
     18 give, in a parametric form, the density profile.
     It is
     clear that the density profile calculated from the above
     approximation depends crucially on the mass growth rate. Hiotelis (2003)
     used a model based on the above assumptions for haloes that grow by accretion
     of matter. This means that the infalling matter is in a form of small haloes (relative
     to the mass of the growing halo) and the procedure can be approximated by a continuous
     infall. In that case, it was shown that  large  growth rate
     of mass (observed in heavy haloes)  leads to inner density profiles that are
      steeper than those of less massive haloes (where the growth rate of mass is
      smaller).\\
      We recall that the density profile of dark matter haloes is
      a very difficult problem contaminated by debates between the results of observations,
       the results of
       numerical simulations and the ones of analytical methods (e.g. Sand et. al 2004 and
       references therein). Especially the inner density profile seems to follow a law of the
       form $\rho (r)\propto r^{-\gamma}$ but the value of  $\gamma$ is not still known.
       Observations estimate a value of $\gamma$ about $0.52$ (Sand et. al 2004), while the results of N-body simulations report
       different values as $1$ (Navarro et al. 1997), or between $1.$ and $1.5$ (Moore et al. 1998).
       Additional results from N-body simulations, as well as from analytical studies, connect
       the value of $\gamma$ with the total mass of the halo (Ricotti 2002, Reed et
       al. 2005,  Hiotelis 2003).  In any case, it is difficult to explain the differences
       appearing in the results of different approximations. For example, regarding N-body
       simulations, some of the results are effects of force resolution. In a typical N-body
       code, such as TREECODE (Hernquist 1987) the force acting on a particle is given by the
       sum of two components: the short-range force that is due to the nearest neighbours
       and the long-range forces  calculated by an expansion of the gravitational potential
       of the entire system. As it can be shown, the value of the average stochastic force in the
       simulation is an order of magnitude greater than that obtained by the theory of stochastic
       forces. Consequently, small fluctuations induced by the small-scale substructure are not
       'seen'. This is the case of cold dark matter models
       in which  the stochastic force generators are substructures
       at least three orders of magnitude smaller in size than the
       protoclusters in which they are embedded (e.g. clusters of
       galaxies). On the other hand, regarding  the differences between the results of various
       approximations, we notice that some of these differences  could be due to the
        presence of the
       baryonic matter.  Especially  in regions, as the central regions of haloes,
        the large density of baryonic matter may affect the distribution of dark
        matter.\\
        Since in this paper we describe in detail
       technical problems related to the construction of reliable merger-trees, we have to note
        that the inside-out model for the formation of haloes described above,
         requires calculations for very small values of $a$, because the central regions
       are formed earlier. Therefore, in order to explore the center of the halo we have to
       go deep in the past of its history. Consequently, the problems relative to the
        calculation of
       mean mass that analyzed at the beginning of this section, become more difficult.
       The density profiles predicted by the above described approximation are under study.

\section{Discussion}
      Analytical models based on the extended PS formalism provide a very useful tool
       for studying the merging histories
      of dark matter haloes.  Comparisons with the results of high resolution
       numerical simulations or
      observations  show that the predictions of  EPS models are satisfactory
      (e.g. Lin et al. 2003,
      Yahagi et al. 2004, Cimatti et al. 2002, Fontana et. al 2004), although
      differences between various approximations  always exist.
      These differences show the need for further improvement of various
      methods.\\
    In this  paper we described in detail the numerical algorithm for the construction
    of merger-trees. The
    construction is based on the mass-weighted probability relation
    that is connected to the  model barrier.  The choice of the barrier is crucial for
    a good agreement between the results obtained by  merger-trees  and those of N-body
    simulations. As it is shown in Yahagi et al. (2004) mass functions resulting
    from the SC model are far from the results of N-body simulations while those
    predicted by non-spherical models are in good agreement.  Our results show
    that the distributions of formation times predicted by the non-spherical models
    are shifted to smaller values. Thus,
    the resulting formation times are closer  to the results of
    N-body simulations than the formation times predicted by the SC model.
    It should be noted that N-body simulations give, in general, smaller formation times
    than that of EPS models (e.g.  Lin et al. 2003). It should also be noted that
    observations indicate a build-up of massive early-type
    galaxies in the early Universe even faster than that expected from
    simulations (Cimatti, 2004).\\
     We have shown that a
     small sample of haloes ($\geq 1 000$) is sufficient for the
     prediction of
   smooth mass growth histories.  These  curves have a functional
    form
    that is in agreement with  those proposed by the N-body results
    of other authors. In particular the set of relations:
    \begin{equation}
    a_c(M_0)=0.24M^{0.12}_{0}
    \end{equation}
    and
    \begin{equation}
     M(a)/M_0=\exp[-2 a_c(M_0) (1/a-1)]
    \end{equation}
     describes very satisfactorily the results of non-spherical models (NR and EC). These
     relations give the  mass $M$  of dark matter haloes at scale
     factor $a$ for the range
     of mass described in the text and for the particular cosmology
     used.
     $M_0$ is the present day mass of the halo in units of
     $10^{12}\mathrm{h^{-1}} M_{\odot}$.\\
     Finally, it should be noted that a large number of questions regarding the formation
      and evolution
     of galaxies remains open. For example, a first step should be the improvement
     of the model barrier so that the results of merger-trees  fit better both
     the mass function (Yahagi et al. 2004)  and the collapse scale factor
     (Lin et al. 2003).
     It should be noted also that structures appear to form earlier in N-body
     simulations and even earlier in real Universe. Additionally,
     predictions relative to the density profile are of interest.
     Some of these issues are currently under study.

\section{Acknowledgements} We are grateful to the anonymous referee for helpful
 and constructive comments and suggestions. N. Hiotelis acknowledges
 the \emph{Empirikion Foundation} for its  financial support and
 Dr M. Vlachogiannis for assistance in manuscript preparation.

\end{document}